\def\full{1}
\def\short{0}
\def\version{\short} 

\documentclass{llncs}
\ifnum\version=\full
\usepackage{fullpage}
\else
\usepackage[margin=1.6in,paper=letterpaper]{geometry}
\fi

\usepackage{amssymb,amsfonts,amsmath,latexsym}
\usepackage{url}
\usepackage{graphicx}

\newcommand{\remove}[1]{}
\newcommand{\snote}[1]{\textsf{[[[]]]}}
\newcommand{\ynote}[1]{\textsf{[[[]]]}}

\newcommand{\secref}[1]{Section~\protect\ref{#1.sec}}

\newcommand{\figref}[1]{Figure~\protect\ref{#1.fig}}
\newcommand{\tblref}[1]{Table~\protect\ref{#1.tbl}}

\newcommand{\eqnref}[1]{Equation~(\protect\ref{#1.eqn})}

\def\eqdef{\stackrel{\mathrm{def}}{=}}

\def\eps{\varepsilon}

\def\HW{\mathsf{Hwt}}
\def\rtr{r_{\mathsf{tr}}}

\def\xor{\oplus}

\def\half{\frac{1}{2}}

\begin{document}

\author{Tzipora Halevi\inst{1} \and Nitesh Saxena\inst{1} \and Shai
Halevi\inst{2}}
\institute{Polytechnic Institute of New York University
\and IBM T. J. Watson Research Center}

\title{Using HB Family of Protocols for Privacy-Preserving Authentication of
RFID Tags in a Population }

\maketitle

\begin{abstract}

In this paper, we propose an HB-like protocol for privacy-preserving
authentication of RFID tags, whereby a tag can remain anonymous and untraceable
to an adversary during the authentication process.  Previous proposals of such
protocols were based on PRF computations. Our protocol can instead be used on
low-cost tags that may be incapable of computing standard PRFs. Moreover, since
the underlying computations in HB protocols are very efficient, our protocol
also reduces reader load compared to PRF-based protocols.

We suggest a tree-based approach that replaces the PRF-based authentication
from prior work with a procedure such as HB+ or HB\#.  We optimize the
tree-traversal stage through usage of a ``light version'' of the underlying
protocol and shared random challenges across all levels of the tree. This
provides significant reduction of the communication resources, resulting
in a privacy-preserving protocol almost as efficient as the underlying
HB+ or HB\#.

\end{abstract}



\setcounter{page}{1}

\vspace{-5ex}
\section{Introduction}
Radio Frequency Identification (RFID) technology is increasingly used
in many aspects of daily life. Low-cost RFID devices have numerous
applications in military, commercial and medical domains.
In most of these applications, an RFID reader must authenticate its
designated tags to prevent tag forgery and counterfeiting. At the same
time, privacy concerns in many of these applications dictate that the
tag's identity is not revealed to an attacker, even while
the tag authenticates itself (and even if the attacker pretends to be a
reader).

RFID tags are typically very low-cost devices and their computation and storage
capabilities are severely constrained. Hence, traditional authentication
protocols based on Pseudo-Random Functions (PRFs) may not be applicable on such
tags.
To this end, Juels and Weis \cite{JW05} suggested using the HB protocol (of
Hopper and Blum \cite{HB01}) for tag authentication.  Several enhancements such
as HB+ \cite{JW05} and HB\# \cite{GRS08} were proposed later, and are currently
the only viable solutions for very low-cost authentication.

Further complicating RFID authentication is the fact that a single
reader must often handle a very large tag population, possibly in the
millions or even billions. Clearly, to use the appropriate secret
keys, the reader must learn the identity of a tag that is trying to
authenticate itself. Although the tag can identify itself by sending
a pre-shared unique identifier, such identifiers can easily be used by
an eavesdropping adversary to compromise the privacy and anonymity of
the tags (and thus also of their bearers). Therefore, a privacy-preserving
solution is needed to allow the reader to identify the tag.

One possible approach to achieve \textit{privacy-preserving authentication}
is for the reader to perform an exhaustive search, by trying to match the
result of the tag being authenticated against all possible tags. However,
this requires a large $O(N)$ amount of computation for a population of
$N$ tags.
To address this scalability problem, Molnar et al. \cite{MW04} proposed
using a ``tree of keys'', assigning to each tag the keys corresponding
to a root-to-leaf path in the tree, and using these keys as seeds for a
PRF; the PRF-based response of a tag is then linked to this path, and
this link is used to identify the tag. This solution reduces the reader
computational complexity of $O(log(N))$.
Also, Cheon et al. \cite{CHT09} proposed a 2D-mesh scheme that enables
the reader to identify the tag with complexity of $O(\sqrt N \log N)$:
The idea is to create two sets of PRF keys, each with $\sqrt{N}$ keys,
and give every tag one key from each of the two sets, such that the
combination of these two keys uniquely determined that tag.
In \cite{BMM08}, Burmester et al. also utilized PRF functions for an
anonymous RFID authentication scheme with constant lookup. In this
approach, the reader
keeps for every tag a constant number of pseudonyms. At the
end of each successful authentication session the tag generates a new
pseudonym, which the reader then uses to identify the tag from its
tag lookup pseudonyms table.

All the above proposals are based on protocols that require the tag to
compute PRF operations and therefore not applicable to very low-cost
tags.
To this end, we consider the HB family of protocols for privacy-preserving
authentication. Adapting HB-like protocols to this setting takes some care,
however. For example, HB protocols are typically designed with very tight
parameters, which yield a non-negligible false accept rate (FAR), i.e., the
probability of an illegitimate tag successfully authenticating to the reader.
In an exhaustive search, the overall false accept rate grows roughly by a
factor of~$N$.  For example, an instance of the HB+ protocol with 80 rounds and
noise probability of 0.25, has an FAR of $4 \times 10^{-6}$. When using this
instance, with an exhaustive search over a population of a million tags, the
overall FAR would be close to one, which is clearly unacceptable for any practical
use.

\subsection{Our Contributions}
In this work we develop a tree-based protocol for privacy-preserving
authentication. Specifically, we study the use of HB-like protocols
(such as HB+ or HB\#).
\ifnum\version=\full
This is useful for two reasons: First, for low-cost tags, not capable
of efficiently performing PRF operations, HB protocols are currently
the only viable solution for authentication. Second, underlying
computations in HB protocols are extremely efficient
which also helps naturally reduce reader load, even when compared to
tree-based protocols for tags capable of performing PRFs.

\fi
Our tree-based protocol has two logical stages. First there is a
\emph{tree traversal stage}, in which the reader tries to identify the
most likely tag to authenticate, and then there is an \emph{authentication
stage}, in which the identity of that ``most likely tag'' is verified.
Unlike exhaustive search, the false-accept rate of this protocol does
not grow with the number of tags in the population, since only one tag
is authenticated in the last stage. We also show that the false-reject
rate grows very slowly, despite the presence of a large number of tags.
We propose several optimizations over naive use of HB+/HB\# in a
tree-based scheme. These optimizations bring the communication of the
tree-based protocol down to not much more than the underlying HB+/HB\#
authentication protocol, and also improve the computation time.

Our work is based on the observation that the function $f_x(A)=Ax+noise$
for a random matrix~$A$ behaves like a (randomized) pseudo-random
function (which was proven in \cite[Lemma~3.2]{APP09} under the
assumption that learning parity with noise is hard). Therefore we
replace the standard PRFs during the tree-traversal stage with this
``HB-like'' function. We observe that this function only needs
tag-generated random challenges during the tree-traversal stage.
(It is only in the authentication stage that a reader-generated
challenge is needed.) Moreover, there is no need to generate a new
challenge in every level, it is perfectly safe to use the same random
challenge in all the levels of the tree, and we can use significantly
smaller parameters during the tree-traversal stage.
The combination of these observations reduces computation and
communication by a factor of~$2\times$ to $4\times$ (In fact our
tree protocol is almost as efficient as the underlying HB+ or HB\#).
We present analytical and simulation results comparing our method with
prior proposals in terms of computation, communication and memory
overheads.

We note that just like all other HB-type protocols, ours is also
vulnerable to man-in-the-middle attacks \cite{GRS05,QOV08}. Also, just
like other tree-based protocols, its privacy guarantees degrade somewhat
when the secret keys of many tags are exposed to the adversary.

\section{Background: HB-type Protocols} \label{hb.sec}
In the original HB protocol by Hopper and Blum \cite{HB01}, the reader
and the tag share a secret $\vec x$ which is a $k$-bit vector (e.g.
$k=256,512$, etc.). The protocol consists of many rounds. In each
round the reader sends a random challenge vector $\vec a$ and the tag
replies with one bit $z=\vec{a}\cdot\vec{x} \oplus \nu$, where $\nu$
is 1 with probability $\eps$ (for some parameter $\eps < 0.5$). After
$r$ such rounds, the tag is accepted if $z$ matches $z'=\vec{a}\cdot
\vec{x}$ at least $r-\tau$ times, where~$\tau$ is some threshold
parameter.

The HB protocol is vulnerable to attacks by an active adversary, who
can send to the tag the same challenge $\vec a$ many times, thereby
eliminating the noise and recovering the secret $\vec x$.  To
counter this attack, the HB+ protocol by Juels and Weis \cite{JW05}
added a second secret $\vec y$ that is shared by the reader and the
tag. In each round of the protocol, a random challenge $\vec b$ is
generated by the tag, a second challenge $\vec a$ is generated by the
reader, and the tag sends to the reader $z=\vec{a}\cdot\vec{x} \oplus
\vec{b}\cdot\vec{y} \oplus\nu$. As before, the tag is accepted after
$r$ rounds if $z$ matches $z'=\vec{a}\cdot\vec{x}\oplus \vec{b}\cdot
\vec{y}$ at least $r-\tau$ times.

It was later shown in \cite{KS05,KS06} that all the rounds can be
carried out in parallel: The tag chooses a matrix B, the reader responds
with a matrix A, the tag chooses a noise vector $\vec \nu$ and replies
$\vec z = A \cdot \vec x \oplus B \cdot \vec y \oplus \vec\nu$, and
the reader checked that $\vec z$ is close enough to $\vec{z'}= A\cdot
\vec x \oplus B\cdot\vec y$. This protocol can withstand an active
attacker that interacts with the tags, but it is still vulnerable to
man-in-the-middle attacks \cite{GRS05,QOV08}.

The HB\# protocol \cite{GRS08} is a variant where the ``types'' of the
secrets and the challenges are swapped: Namely, the tag and reader share
secrets $X$ and $Y$ which are $r\times k$ matrices, the challenges are
$r$-vectors $\vec{a}$ and~$\vec{b}$, and the reply that the tag computes
is $\vec{z}=\vec{a}\cdot X \oplus \vec{b}\cdot Y \oplus\vec\nu$.
To alleviate the additional storage requirement, HB\# proposes using
Toeplitz matrices for $X,Y$.
Another claimed benefit for swapping the types is that HB\# can resist some
limited form of man-in-the-middle attacks, however, not all \cite{QOV08}.

\begin{center}
\begin{figure}
\setlength{\unitlength}{.25in}
\begin{picture}(23,6)(0,0)
\linethickness{0.5pt}
%
\put(3,6){\makebox(0,0){\underline{Reader $(\vec{x},\vec{y})$}}}
\put(16.5,6){\makebox(0,0){\underline{Tag $(\vec{x},\vec{y})$}}}

\put(3,5){\makebox(0,0){$\vec a \in n \{0,1\}^k$}}
\put(16.5,5){\makebox(0,0){$\vec b \in n \{0,1\}^k$}}

\put(10,5){\makebox(0,0){$\vec b$ - challenge}}
\put(13.5,4.75){\vector(-1,0){8}}

\put(10,4){\makebox(0,0){$\vec a$ - challenge}}
\put(5.5,3.75){\vector(1,0){8}}

\put(16.5,3.1){\makebox(0,0){$\nu \in \{0,1 | Prob (v=1) = \eps\} $}}

\put(16.75,2.5){\makebox(0,0)
{$z=\vec{a}\cdot\vec{x} \oplus \vec{b}\cdot\vec{y}\oplus \nu$}}

\put(10,2){\makebox(0,0){z - response}}

\put(13.5,1.75){\vector(-1,0){8}}

\put(3,1.25){\makebox(0,0){$z'=\vec{a}\cdot\vec{x}\oplus\vec{b}\cdot\vec{y}$}}

\put(3,0.5){\makebox(0,0){Compare $z \stackrel{?}{=} z'$}}

\framebox(20,6.5)[-6,-1]{}
\end{picture}
\caption{One round of the HB+ Protocol}
\end{figure}
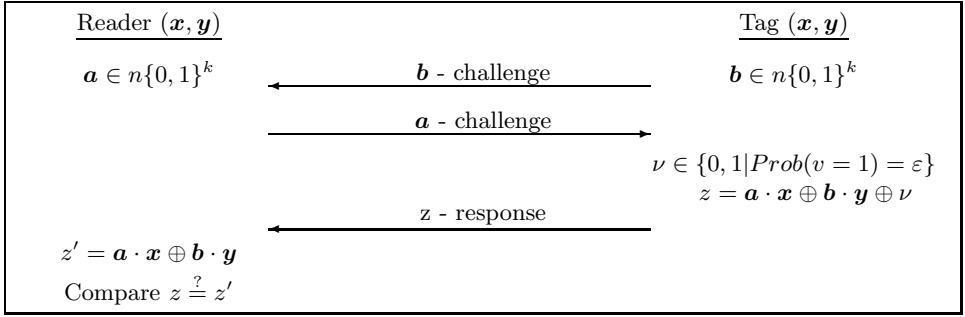
\vspace{-6ex}
\end{center}

\subsection{Security}
The security of the HB series of protocols relies on the computational
hardness of the problem of \emph{Learning from Parity with Noise} (LPN)
\cite{JW05}. Roughly, the LPN problem is to find a secret $k$-vector
$\vec x$ given a random binary $r\times k$ matrix $A$ and a vector
$\vec z=A \cdot \vec{x}\oplus\vec{\nu}$, where $\nu$ is a noise vector
in which each entry is set to~1 with probability $\eps$ (for a parameter
$0 < \eps < \frac12$).

The hardness of this problem (and thus the security level of the
protocols) depends on the choice of the parameters $\eps$ and~$k$.
The best algorithm for learning parity with noise is due to Blum,
Kalai and Wasserman \cite{BKW03}, with some later optimizations,
notably by Levieil and Fouque \cite{LF06}. Based on the performance of
the BKW algorithm, Juels and Weis suggested in \cite{JW05} that
setting $k=224$ and $\eps\in[\frac18,\frac14]$ yields an 80-bit
security level. In \cite{LF06}, Levieil and Fouque claimed, however,
that these parameters yield only a 55-bit security level, and that
with $\eps=\frac14$ one needs to use $k=512$. (They also
noted that only one of the secrets $\vec{x},\vec{y}$ in HB+ needs to
be this long, while the other can safely be only 80-bit long.)

The other parameters of interest in HB-type protocols are the number
of rounds%
\footnote{The name ``number of rounds'' refers to a sequential
  protocol where a one-bit~$z$ is returned in every round. For the
  parallel version, this parameter is the reply-size.
}\
and the acceptance threshold. Juels and Weis suggested in \cite{JW05}
to use number of rounds $r\in[40,80]$ and set the threshold at $\tau=\eps
r$. This gives a relatively high false reject rate (FRR) of 0.44, and
when used with error probability $\eps\in[\frac18,\frac14]$ it gives
false-accept rates (FAR) between $10^{-3}$ and $10^{-9}$.
(The HB\# proposal in \cite{GRS08} suggested other parameters with a
much lower FRR and FAR, e.g., FAR~$\le 2^{-80}$ and FRR~$\approx
2^{-45}$.)

\section{Private Authentication with an HB-like protocol}\label{multi-HB.sec}
\ifnum\version=\full
As discussed in the introduction, previous algorithms for
privacy-preserving authentication rely on the ability of the tag
to perform PRF computation. We now propose a method that only
requires HB-like computation, and therefore may be more suitable
for very low-cost tags that can not perform PRF calculations. Moreover,
the use of HB-like protocol also reduces the reader-side computation
(since at reader's end also computing HB-like operations is cheaper than a
PRF such as AES).

\fi

Recall that in our setting, we have a system with many tags under one
administrative domain. Similarly to \cite{MW04}, we arrange all the
tags in a tree structure and use this structure to find the right tag
to authenticate. However, we use an HB-like procedure to implement
both the tree search and the tag authentication as opposed to the
PRF-based solution in \cite{MW04}.

A naive approach would simply perform the underlying authentication
protocol at every level of the tree. Namely, at every level one can
run HB+ (or HB\#) and the reader then performs an exhaustive search
over the children of the current node, selecting the one for which the
authentication was successful.%
\footnote{In the rest of this paper, we concentrate only on the case
  of HB+. The treatment of HB\# is completely analogous.
}\
A closer inspection reveals that this naive approach can be
significantly improved:
\begin{itemize}
\item
First, there is no reason to use challenges from the reader while
traversing the tree. Indeed, the challenge from the reader is needed
for secure authentication, but $\textit{not}$ for the identification
of ``the right tag to authenticate''.

We therefore suggest to logically split the protocol into a
\emph{tree-traversal stage}, where the reader finds the right tag to
authenticate, and an \emph{authentication stage}, where the identity
of that tag is verified. In the authentication stage, we just use the
underlying HB+ protocol unchanged, but in the tree-traversal
stage we only use random challenges from the tag. This optimization
cuts the computation time and storage requirements for both the tag
and the reader.

(We note, however, that in the asymmetric setting suggested in
\cite{LF06}, where one secret is longer than the other, we need to
store a long secret in every internal node in the tree. For example,
using the 512/80 asymmetric setting, this optimization only saves
80 of the combined $512+80=592$ bits, which is $\approx 13\%$.)
\smallskip

\item
More importantly, we observe that we can share the same challenges
across all the levels. Namely, instead of the tag sending a different
challenge for every level, we send only one challenge and use it
at all the levels. This means that we only have communication of
two challenges throughout the tree-based protocol (just as in the
underlying HB+), and the only additional communication is an
$r$-bit response vector for every level.
\end{itemize}

In more details, the reader in our system maintains a tree structure
with the tags at the leaves, and each internal node $n$ in the tree is
associated with a secret key $\vec{y_n}$, which is a binary vector of
length~$k_y$. A tag~$t$, associated with a leaf in the tree, is given
all the secret keys that are associated with nodes on the path from
the root to that leaf. In addition the tag is also given two unique
secret keys $\vec{x}_t$ and $\vec{y}_t$ (of length $k_x$ and~$k_y$,
respectively) that are used for authentication. The notations that we
use in the description below are summarized in \tblref{notations}.

\vspace{-2ex}

\begin{table}[ht]
\caption{Notation table \label{notations.tbl}}
\centering
\begin{tabular}{|c| l|} \hline
$d$ & ~depth of the tree \\
$\beta$ &  ~branching factor of the tree  \\
~$k_x,k_y$ &  ~length of the secrets in the system $(k_x=80, k_y\in[224,512])$\\
$r$ & ~size of the ``response'' messages (usually $r\in[80,128]$) \\
$\eps$ & ~noise level ($\eps=0.25$) \\
$\tau$ & ~acceptance threshold (usually $\tau \in [20,40]$)\\
\hline
\end{tabular}
\end{table}

\vspace{-3ex}

\subsection{System Setup}
The system is setup with some upper bound~$N$ on the number of tags that
it can support, and the parameters $r,d,\beta$ are derived from this
upper bound. (See \secref{params} for a discussion of how to determine
these parameters.) Once determined, the parameters $d,\beta$ define a
tree with $\beta^d\ge N$ leaves, which will be associated with the tags
in the system. Also, a ``master secret'' is chosen, which will be used
to determine all the other secrets in the system as described next.

\subsection{Tag Registration}
When a tag is registered into the system, it is associated with a random
available leaf in the tree. This can be done either by the system
administrator remembering the positions of all the tags in the tree and
choosing at random one leaf that is not yet used, or by choosing a
random permutation~$\pi$ over the domain $[1,\beta^d]$ and assigning the
$i$'th tag in the system to leaf $\pi(i)$.
\footnote{An efficient method for implementing such a random permutation
was recently proposed by Morris et al. \cite{MRS09}.}\

Let $n_0,n_1,\ldots,n_d$ be the path in the tree from the root (denoted
$n_0$) to the leaf that is associated with the new tag (denoted~$n_d$).
For each node $n_i$ (except the root), the tag is given the $k_y$-bit
secret $\vec{y}_{n_i}$ that is associated with that node. (This secret
can be derived from the master secret, say by setting $\vec{y}_{n_i}=
PRF_{ms}(n_i)$ for some pseudo-random function PRF that is keyed by
the master secret~$ms$.)
The tag also gets two additional keys $\vec x_t$ and $\vec y_t$ of
size~$k$ (that can similarly be derived from the master secret).
After registration, the tag will hold the keys $\vec{y}_{n_1},\ldots,
\vec{y}_{n_d}, \vec{x}_t, \vec{y}_t$.

\subsection{Tree-Traversal Stage}
At the start of the authentication process, the tag needs to be
identified so the correct keys can be used for the authentication. To
this end, the tag chooses a $r\times k_y$ random challenge matrix~$B$,
and noise vectors $\nu_i$ for every level~$i$ in the tree. The tag computes
$\vec{z_i}=B\cdot\vec{y}_{n_i} \oplus \vec{\nu_i}$ and sends to the
reader the matrix~$B$ and all the $\vec{z}_i$'s.

Starting from the root, the reader then goes down node by node, using
the $\vec{z}_i$'s to decide what child of the current node it needs to
use next. Specifically, for every child~$c$ of the root it computes
$\vec{z}_c = B\cdot \vec{y}_c$ , where $\vec{y}_c$ is the secret
associated with that child. Then, the reader descends into the
child~$c$ for which $\vec{z}_c$ is closest to the response vector
$\vec{z}_1$ received from the tag. Similarly, after descending into an
internal node~$n_i$ at level~$i$, the reader computes $\vec{z}_c =
B\cdot \vec{y}_c$ for every child~$c$ of $n_i$, and descends into the
child~$c$ for which $\vec{z}_c$ is closest to $\vec{z}_i$ that was
received from the tag. (Throughout this process, if two children are
equally close to $\vec{z}_i$, one is chosen arbitrarily as the next
node).

After using all the $\vec{z}_i$'s, the reader arrives at one of the
leaves of the tree, and then it uses the secrets $\vec{x},\vec{y}$
that are associated with the tag at that leaf in the authentication
stage, as described next.

\subsection{Authentication Stage}
At the end of the Tree-traversal stage, the reader arrives at a leaf that
it considers to be the most likely to be the one associated with the
correct tag. At this stage, we need to run the authentication protocol
to confirm that the tag is valid. Here we just use the parallel HB+
protocol. Specifically, the reader sends a random challenge matrix~$A$,
the tag chooses a noise vector $\vec{\nu}$, and reply with $\vec{z}=
A\cdot\vec{x}_t \oplus B\cdot\vec{y}_t\oplus \nu$ (using the two last
secret keys $\vec{x}_t,\vec{y}_t$ that it holds). The reader recovers
the keys $\vec{x},\vec{y}$ that are associated with the tag in this
most likely leaf, computes $\vec{z'}=A\cdot\vec{x} \oplus B\cdot
\vec{y}$, and checks that the Hamming distance between $\vec{z}$
and~$\vec{z'}$ is below the threshold~$\tau$. It accepts the tag if
$\vec{z}$ and~$\vec{z'}$ are close enough, and rejects it otherwise.

\section{Security Analysis} \label{security.sec}
Recall that we have two security goals in our model, namely privacy
and authentication. Our attack model allows the adversary to eavesdrop
on the communication between tags and the reader, and also to
communicate directly with the tag and the reader, but NOT to modify
messages that are sent between them. In other words, we consider an
active adversary, but explicitly disregard man-in-the-middle attacks
(since all HB-type protocols are insecure against them
\cite{GRS05,QOV08}).

\subsection{Authentication}
Our model for authentication is essentially the usual
\textsc{det} model \cite{JW05,KS05,GRS08}, where in our case we have
one target tag that the attacker is trying to impersonate, and all the
other tags in the system are considered to be adversarial. In more
details, the attacker first gets the secrets of all the tags in the
system but one, then it can interact with the remaining tag for~$q$
times, and finally it interacts with the reader and tries to impersonate
that remaining tag.

Since we just run the original HB+ protocol at the authentication
stage, then trivially our protocol is as resilient against impersonation
as HB+. (Formally, there is a trivial tight reduction from breaking
our protocol to breaking the original HB+.)

\subsection{Privacy}
Our privacy model meant to capture an attacker who can interact with
many tags, and who tries to link different authentication sessions.
For example, consider an attacker that roams through a library
talking to tags and recording the books that these tags are embedded
in. Later the attacker detects some tags in the book-bag of a random
person on the street, and it tries to recognize these tags (thereby
recognizing the books that this person checked out).

A (somewhat simplistic) formalization of this concern has the adversary
talking to two tags in the system, which are chosen as follows: First
the adversary chooses two arbitrary distinct tags in the system, and
then a fair coin is tossed. If it comes out \textsf{head} then the two
tags are the ones chosen by the adversary, and if it comes out
\textsf{tail} then the two tags that the adversary talks to are
actually just the first tag that it chose. The adversary now interacts
with the ``two tags'' for a total of $q$~sessions, at the end of which
is needs to guess the outcome of the coin toss.

We note that this is not the only possible definition for privacy in
this setting (and maybe also not the most natural one). For example,
this definition does not consider the possibility of the adversary
compromising some tags and learning their secrets. Still, we think
that this model is a reasonably meaningful one.

The security of our protocol in this model follows easily from the
fact that the function $f_{\vec{y}}(B)=B\cdot\vec{y}+\mathrm{noise}$
is a pseudorandom function when~$B$ is random, and this fact was
proven in \cite[Lemma~3.2]{APP09} to follows from the hardness of
LPN. Indeed, it is easy to see that if we replace this function in
our protocol with a truly random function then the adversary would
have zero advantage in guessing the outcome of the coin-toss in the
experiment from above. (The fact that we can share the same challenge
matrix~$B$ for all levels of the tree follow from an easy hybrid
argument.)

We finally note that our protocol, like all tree-based protocols, is
somewhat vulnerable to tag-compromise. If an attacker manages to get
the secrets of a specific tag, it can use them to recognize the
tree-traversal replies of neighbor tags in the tree (and in particular
it can distinguish neighbors of the compromised tags from non-neighbors).
Devising mechanisms to mitigate the effect of tag compromise is an
interesting open problem.

\section{Parameters and Performance}\label{analysis.sec} \label{errors.sec}
Below we analyze the false reject and false accept rates for a fixed
set of parameters $\beta,d,r$, and then use this analysis to set these
parameters (as a function of the total number of tags~$N$).

\subsection{False-Accept Rate}
We begin by observing that the false-accept rate of our protocol is
exactly the same as that of the underlying HB+ that is used in the
authentication stage. To see why, observe that the tree-traversal stage
of the protocol always results in the reader identifying one leaf of
the tree as the most likely to correspond to the right tag, and then
the underlying HB+ is run against the keys of the tag in this most
likely leaf.

This is in sharp contrast to the situation for the trivial linear search
routine that was discussed in the introduction. In that procedure, the
reader would try to check the authentication against the keys of all the
tags in the system, so the false-accept rate would roughly be multiplied
by the number of tags. This does not happen in our algorithm, since the
reader only tries to authenticate against the keys of just one tag.

In other words, even giving the adversary ``for free'' the ability to
steer the tree-traversal stage to any tag of its choice, it would still have
to be able to authenticate a tag without knowing its keys in order to
induce a false accept event.

\subsection{False-Reject Rate}
We note, however, that the false reject rate of our algorithm will be
higher than the standard HB+. This is because the tree-traversal
stage may identify wrong leaf, in
which case the authentication stage will almost surely reject. We
therefore need to analyze the probability that the tree-traversal stage
identifies the wrong leaf.

\subsubsection{Taking a false branch, the binary case.}
We start by considering a binary tree (i.e., $\beta=2$) and analyzing
the probability that the wrong branch is chosen at one step of the
tree-traversal stage. For the binary tree, if the algorithm is currently in
a node at level~$\ell-1$ which is on the path to the true tag, then
the likelihood of choosing the wrong child for level~$\ell$ is exactly
the probability that the non-matching key of that ``false child''
generates a vector $\vec{z}_f$ that is closer to the response vector
$\vec{z}_{\ell}$ than the matching key of the ``true child''.

Below we denote by $\HW$ the Hamming weight of a binary vector. We denote
the ``true child'' of the current node by~$t$ (i.e., the child on the
path to the true tag that is trying to authenticate itself), and denote
the other child of that node by~$f$ (for a ``false child''). With the
keys of these children denoted $\vec{y}_t$ and $\vec{y}_f$,
respectively, and the tag sending a challenge matrix~$B$, recall that
the reader computes $\vec{z}_f=B\cdot\vec{y}_f$ and
$\vec{z}_t=B\cdot\vec{y}_t$, and compares these two vectors against the
vector $\vec{z}_{\ell}$ that was sent by the tag.

The probability of descending into the wrong child~$f$ is bounded by the
probability that $\vec{z}_f$ is closer to $\vec{z}_{\ell}$ than
$\vec{z}_t$, namely $\Pr[\HW(\vec{z}_f\xor\vec{z}_{\ell}) <
\HW(\vec{z}_t\xor\vec{z}_{\ell})]$. We denote by $P_t(i)$ the
probability that $\vec{z}_t$, $\vec{z}_{\ell}$ differ by exactly $i$
positions, and similarly by $P_f(i)$ we denote the probability that
$\vec{z}_f$, $\vec{z}_{\ell}$ differ by exactly $i$ positions. Namely,

\vspace{-2ex}

\begin{eqnarray*}
P_t(i) &\eqdef& \Pr[\HW(\vec{z}_t\xor\vec{z}_{\ell})=i]
= {r \choose i}\eps^i (1-\eps)^{r-i}
~~~(\mbox{where we use }\eps=1/4)\\
P_f(i) &\eqdef& \Pr[\HW(\vec{z}_f\xor\vec{z}_{\ell})=i] = {r \choose
i}/2^{r}
\end{eqnarray*}

Then we bound the probability of taking a false branch by
\begin{equation}
\Pr\left[\HW(\vec{z}_f\xor\vec{z}_{\ell})
< \HW(\vec{z}_t\xor\vec{z}_{\ell})\right]
~=~ \sum_{i=1}^r P_t(i) \cdot \left(\sum_{j=0}^{i-1} P_f(j)\right)
\label{binarycase.eqn}
\end{equation}

\def\erfc{\mathsf{erfc}}
We can approximate the above expression as follows: Let $\Delta_{ft}$
be the difference between the Hamming weight of
$\vec{z}_t\xor\vec{z}_{\ell}$
and $\vec{z}_f\xor\vec{z}_{\ell}$,
\[
\Delta_{ft} \eqdef \HW(\vec{z}_f\xor\vec{z}_{\ell})
- \HW(\vec{z}_t\xor\vec{z}_{\ell})
\]
The random variable $\Delta_{ft}$ is the sum of~$r$ independent and identically distribution
random variables, one for every position in the response vector (and
each a random variable over $\{-1,0,1\}$). The probability of taking a
false branch is then bounded by $\Pr[\Delta_{ft}<0]$, and by the law
of large numbers we can approximate $\Delta_{ft}$ by a Normal random
variable with the same mean and variance.
Recalling that $\HW(\vec{z}_t\xor\vec{z}_{\ell})$ has mean $\mu_t=
\frac{r}4$ and variance $\sigma^2_t=\frac{3r}{16}$ and that
$\HW(\vec{z}_f\xor\vec{z}_{\ell})$ has mean $\mu_f=\frac{r}2$ and
variance $\sigma^2_f=\frac{r}{4}$, we get that $\Delta_{ft}$ has
mean $\mu=\mu_f-\mu_t=\frac{r}4$ and variance $\sigma^2=
\sigma^2_t+\sigma^2_f=\frac{7r}{16}$. Hence we have
\[
\Pr[\mbox{false branch}] ~\le~ \Pr[\Delta_{ft}<0]
~\approx~ \erfc\left(\frac{r/4}{\sqrt{7r/8}})\right)
~=~ \erfc\left(\sqrt{r/14}\right)
\]

For example, if we choose $r=80$, then we estimate the probability of
choosing a false branch as $\Pr[\mbox{false branch}] \approx 6.97 \cdot
10^{-4}$, while a choice of $r=144$ gives $\Pr[\mbox{false branch}]
\approx 5.38 \cdot 10^{-6}$.

\subsubsection{Taking a false branch, the general case.}
For a larger branching factor ($\beta>2$), we can still use an equation
similar to \eqnref{binarycase} for the probability of choosing a false
branch, but estimating it becomes harder. Specifically, we replace the
probability that one false tag has less errors than the true tag by the
probability that at least one of the false tags has less errors:

\begin{equation}
\Pr[\mbox{false branch}](\beta)~=~ \sum_{i=1}^r P_t(i) \cdot \left[1 -
\left(1-\sum_{j=0}^{i-1} P_f(j)\right)^{\beta}\right]
\label{beta_g2.eqn}
\end{equation}
However, estimating the last expression is harder, since we need to
bound the probability that the minimum of several random variables is
below zero (and moreover these random variables are highly correlated).

In lieu of an analytical bound, we therefore evaluated the expression
from \eqnref{beta_g2} explicitly, and plotted the results in
\figref{flase_branch}. Specifically, fixing the target false-branch
probability to some constant (either 0.1 or 0.01), we plot for every
branching factor from $\beta=2$ to $\beta=10^4$ the smallest
response-size~$r$ for which the false-branch probability is below that
target. (For example, with branching factor of $\beta=1000$ we need
$r\approx 80$ to get false-branch probability of 0.1.) As expected,
for a constant probability of false branch, the response-size~$r$
grows linearly with the log of the branching factor.

\begin{figure}
\begin{center}
\includegraphics[height=2in]{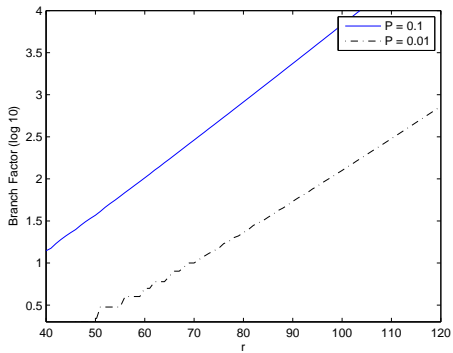}
\caption{Branching factor~($\beta$) vs. response-length~($r$), for
false-branch probabilities of 0.1 and 0.01.
\label{flase_branch.fig}}
\end{center}
\end{figure}

\subsubsection{The overall false reject rate.}
Recall that a valid tag can be falsely rejected either due to
the algorithm choosing a false branch at some point during the
tree-traversal stages, or due to a false reject in the authentication
stage. The probability of failing the authentication stage is equal to
the probability that the error vector has more than $\tau$ ones, namely
the false-reject rate of the authentication stage is
\begin{equation}
FRR(\mbox{auth}) \eqdef 
  \sum_{i=t+1}^{r} \eps^i (1-\eps)^{r-i}{r \choose i}
\end{equation}
and the combined false-reject rate of the whole protocol is
\begin{equation}
FRR(\mbox{Tree-}HB+)
~\approx~ d\cdot\Pr[\mbox{false branch}] ~+~ FRR(\mbox{auth})
\end{equation}

As mentioned in \cite{KS05}, the false reject rate can be reduced if
the tag checks the noise vector $\nu$ and only use it if it has
at most $\tau$ one-bits.


\subsection{Computation, Storage and Communication}
\label{ComAnlsis.sec}
The computation of the reader during the tree-traversal stage consists
of $d$ levels, wherein each level the reader computes $\beta$ vectors
$\vec{z}_c$ (one for every child $c$ of the current node) and compares
them to the response-vector $\vec{z}_i$ that the tag sent. Then, the
computation during the authentication stage consists of a single
execution of parallel-HB+. Recalling that every vector
$\vec{z}_c$ during the tree-traversal stage takes one matrix-vector
product to compute and the authentication stage takes two more
matrix-vector products (one of which with a smaller $k_x\times r$
matrix), we get that the total computation on the reader side is
between $d\beta+1$ and $d\beta+2$ matrix-vector multiplications.

On the tag side, the tag computes one matrix-vector product for every
level of the tree in the tree-traversal stage, and two more in the
authentication stage (again, one of which with a smaller matrix), for
a total of between $d+1$ and $d+2$ products.

As for communication, the tag sends the challenge matrix~$B$ for
parallel-HB+ and the response vectors $\vec{z}_i$, it receives the
challenge matrix $A$ and then sends~$\vec z$. Hence the total
communication (in both directions) is $r(k_x+k_y+d)$.%
\footnote{We note that just like for HB\#, it is likely that here
  too we can reduce the communication to $r(d+2)+k_x+k_y$ by using
  Toeplitz matrices for the challenges $A,B$.
}\
Finally, the storage requirement on the tag is $k_y(d+1)+k_x$ bits.

An important optimization for our protocol is to choose different
values for the response-length~$r$ in the two stages, so as to get
$d\cdot\Pr[\mbox{false-branch}]\approx FRR(\mbox{auth})$. If we
denote by $\rtr$ the response-length
in the tree-traversal stage and by $r$ the response-length in the
authentication stage, we get total communication of
$r(k_x+k_y)+\rtr d$. The effect on computation is even larger:
since we typically have $\rtr<r/2$, the total computation is
decreased by about a factor of two.

\subsection{Iterating the Protocol} \label{iterate.sec}

We recall that an easy (and cheap) way of reducing the false-reject
rate is to iterate the protocol several times. Specifically, given a
protocol $\Pi$ with FRR of~$\gamma$, FAR of~$\delta$, and complexity
$C$, we can get from it a protocol~$\Pi'$ with smaller FRR as follows:
we first run $\Pi$ once, and if the authentication fails then we run
it again. Clearly, $\Pi'$ has FRR of~$\gamma^2$ and FAR of $2\delta$,
and we note that its expected complexity is only $C(1+\gamma)$ (since
with probability $1-\gamma$ we will only run it once).

Similarly, by repeating the original protocol $\Pi$ upto $s$ times, we
obtain a protocol $\Pi^s$ with FRR of $\gamma^s$, FAR of $s\delta$,
and for large~$s$ the expected complexity will approach $C/(1-\gamma)$.
This means that even protocols with very large false-reject rate (such
as the original HB+ that has FRR$\approx 0.44$) are meaningful, in that
we can transform them cheaply to protocols with adequately low FRR.

\subsection{Protocol Comparison}
In \tblref{protcomp.1} we compare the trivial exhaustive search with
HB+ (ES), our tree-based privacy-preserving HB+ protocol, and the
tree-based PRF protocol developed in \cite{MW04}.

Obviously, the
calculation on the reader side takes $O(N)\cdot C_{HB+}$ for
exhaustive search (where $C_{HB+}$ is the number of calculations for
the HB+ authentication protocol), while for the tree-based protocols,
this is reduced to logarithmic in~$N$. We also list in that table
the storage requirements, the communication, and the corresponding FRR
and FAR. (We note that in the exhaustive search case, if the FAR is
too high then the FRR is not meaningful as the probability that a
valid tag will be identified and accepted as a different tag is very
high.)

For this table we used the values of $\beta=1000$ for the branching
factor of the tree and $d=2$ for its depth (for a total population of
$N=10^6$ tags). For HB+ we used $r=80$ for the response length, $\eps=
\frac14$ for the error rate and $\tau=\eps r=20$ for the acceptance
threshold, and for the size of the keys we used the ``low security''
values pf $k_x=80$ and $ky=256$ (corresponding to the requirement
of $2^{55}$ memory to break LPN). For the PRF we used key size and
output size of 128 bits (e.g., for AES-128).

\begin{table}
\caption{Protocol comparison for a population of $N=10^6$ tags.}
\centering 
\begin{tabular}{| c | c | c | c | c | c | } \hline
\textbf{Method} & \textbf{Reader Computation} & \textbf{Communication} & \textbf{Tag Memory} & \textbf{FAR} & \textbf{FRR} \\
[0.5ex] 
\hline\hline 
ES HB+ & $10^6 \cdot C_{HB+}$ & 26960 & 336 & 0.98 & 0.44
\\ \hline
Tree HB+ &  $2000 \cdot C_{HB+}$ & 27120 & 848 & $4 \cdot 10^{-6}$ & 0.6
\\ \hline
Tree PRF & $ 2000 \cdot C_{PRF}$ & 1024 & 256 & 0 & 0 \\
\hline 
\end{tabular}
\label{protcomp.1.tbl} 
\end{table}

\vspace{-5ex}
\subsection{Choosing the Parameters} \label{params.sec}
With the analysis from above, we now illustrate how to set the parameters of our scheme for a given tag population and security goals. The results of this sections are summarized in \tblref{concretePrms}. Specifically, suppose that we want to setup the scheme to achieve the following parameters:
\begin{itemize}
\item Tag population of upto $N=10^6$ tags,
\item False-reject rate of $10^{-4}$ and False-accept rate of $~10^{-8}$,
\item Security against attacks that work in space of upto $2^{65}$ bytes.
\end{itemize}
For these settings, we investigate the parameters that we get by working with noise rates of $\eps=0.125$ or $\eps=0.25$, using tree of depth either $d=2$ or $d=3$ (and thus $\beta=1000$ or $\beta=100$, respectively).

\subsubsection{The key-length $k_x,k_y$.}
Extrapolating from the parameters in \cite[Sec~5.2]{LF06}, for this level of security we need $k_y\approx 440$ for noise level $\eps=0.125$ and $k_y\approx 330$ for noise level $\eps=0.25$ (and in either case we can get by with $k_x=80$).

\subsubsection{The response-length~$r$ and threshold~$\tau$.}
As mentioned above, we can always get good false-reject rates by repeating the protocol a few times, as long as we start from a protocol that has low enough false-accept rates. Stipulating that repeating the protocol upto four times is reasonable, we thus begin by concentrating on the authentication phase, setting our target for a single run at (say) FAR$\approx 10^{-9}$ and FRR$\approx 0.05$.

To get these parameters for the case $\eps=0.25$ we can use response-length $r=212$ and threshold $\tau=63(=212/4~+10)$, which yield FAR$=1.62E-9$ and FRR$=0.038$. Similarly, for the $\eps=0125$ case we can use $r=86$ and $\tau=16(=86/8~+5.25)$, which yield FAR$=1.6E-9$ and FRR$=0.036$.

To optimize performance, it is beneficial to set a different response-length for the tree-traversal phase, so as to get $d\cdot\Pr[\mbox{false branch}] \approx FRR(\mbox{auth})$. Below we denote the response-length for the tree-traversal by $\rtr$.
For $\eps=0.25,\beta=1000$ we need  $\rtr=102$ (yielding $\Pr[\mbox{false branch}]=0.025$),
for $\eps=0.25,\beta=100$ we need $\rtr=83$ ($\Pr[\mbox{false branch}]=0.0167$),
for $\eps=0.125,\beta=1000$ we need $\rtr=40$ ($\Pr[\mbox{false branch}]=0.0215$), and
for $\eps=0.25,\beta=100$ we need $\rtr=32$ ($\Pr[\mbox{false branch}]=0.0146$).

\subsubsection{The resulting parameters.}
The choices above result in four sets of parameters, depending on the values of the noise rate $\eps\in\{0.125,0.25\}$ and the depth of the tree $d\in\{2,3\}$ (corresponding to branching factors $\beta\in\{1000,100\}$, respectively). In either case, running the protocol once induces total communication $r(k_x+k_y)+ \rtr d$, computation of $d k_y\rtr+(k_x+k_y)r$ bit operations on the tag and $\beta d k_y\rtr+(k_x+k_y)r$ bit operations on the reader, and storage requirements of $k_x+(d+1)k_y$ for the tags.
Running the protocol once with these parameters result in a false-accept rate of roughly 0.1, which is too high for applications, so we repeat it upto four times. As we explained in \secref{iterate}, this only increases the expected complexity of the protocol by roughly $10\%$.
The result (when running the protocol upto four times) are summarized in \tblref{concretePrms}.

\begin{table}
\caption{Concrete parameters for population of $N=10^6$ tags, running the protocol upto four times to reduce the false-accept rates.}
\centerline{\begin{tabular}{|r|c|c|c|c|c|c|c|c|c|c|c|c|c|}
\hline
$\eps$~~ & ~~$d$~~ & $\beta$ & ~$k_x$ & $k_y$ & $r$ & ~$\rtr$~
     & $C_{\mathsf{rdr}}$ & $C_{\mathsf{tag}}$ & \textsf{comm} & \textsf{mem} & FRR & FAR \\  [0.2ex]
\hline\hline
0.25 & 2 & 1000 & 80 & 330 & 212 & 102 & \;$7.49E+7$ & \;$1.71E+5$
     & 96804 & 740 & \;$6.0E-5$ & \;$6.5E-9$
\\ \hline
0.25 & 3 & 100  & 80 & 330 & 212 & 83 & \;$9.23E+6$ & \;$1.88E+5$
     & 96854 & 1400 & \;$6.0E-5$ & \;$6.5E-9$
\\ \hline
0.125 & 2 & 1000 & 80& 440 & 86 & 40 & \;$3.92E+7$ & \;$8.88E+4$
     & 49778 & 1400 & \;$3.9E-5$ & \;$6.4E-9$
\\ \hline
0.125 & 3 & 100 & 80 & 400 & 86 & 32 & \;$4.74E+6$ & \;$9.66E+4$
     & 49796 & 1840 & \;$4.1E-5$ & \;$6.4E-9$
\\ \hline
\end{tabular}
}
\smallskip

Legend:
$\eps$- error-rate, $d$ - depth, $\beta$ -branching factor,
$k_x,k_y$ - key lengths, $r$ - response length in auth. stage,
$\rtr$ - response length in tree stage,
$C_{\mathsf{rdr}}$ - expected reader computation,
$C_{\mathsf{tag}}$ - expected tag computation,
\textsf{comm} - expected total communication,
\textsf{mem} - tag memory requirements.
\label{concretePrms.tbl}
\end{table}

It can be seen from \tblref{concretePrms} that most of the relevant
parameters are improved significantly when moving to a lower-error
regime (i.e., from $\eps=0.25$ to $\eps=0.125$), and also when moving
to deeper trees (i.e., from $d=2$ to $d=3$). The only limiting parameter
is the memory requirement at the tag, which increases for deeper trees
and smaller error rates. Therefore, for any particular application, one
should use as deep a tree and as small an error as can be realized
subject to the memory available at the tags.

We again note that the communication complexity of the protocol (which
is rather large) can probably be significantly reduced by using Toeplitz
matrices for the challenges $A,B$.%
\footnote{Another alternative is to replace $A\vec{x}+B\vec{y}$ with
  $\vec{a}\cdot\vec{x}+\vec{b}\cdot\vec{y}$ over a finite field.
}\
(In the examples above, this optimization will reduce the communication
complexity from the current range of 50000-100000 bits to only about
1000 bits.)
Another ``big win'' will be to be able derive the keys $\vec{y}$
pseudo-randomly manner from shorter secrets (e.g., using a cheap PRG
such as the shrinking generator \cite{CKM93}).

\section{Simulation} \label{simulation.sec}
To estimate the computational overhead incurred at the reader-side
when using different methods, we performed a simulation. Our
simulation was done using Matlab. The computer used for the simulation
is an IBM thinkpad, with Intel CPU, T2400, 1.83 GHz and 1.99 GB RAM.

The test was run for a million tags. We tested two cases for our
Tree-HB+ protocol: A two-level tree with 1000 nodes in each level, and
a three-level tree with 100 nodes at each level, all with the
parameters that were proposed in the original HB+ protocol (i.e.,
error-rate of $\eps=0.25$, using $k_x=k_y=224$).
Our results show that
for $r=96$, the exhaustive search average run-time using HB+ was 110
seconds. For the tree-based search (2 levels), we get an average run
time of 0.1307 seconds, and for 3 levels we get run time of
0.019. This ratio is compatible with our theoretical estimates, as the
exhaustive search will take $O(N)$ (e.g., $\approx 10^6$) while the
tree based search only takes $O(d\log N)$ (e.g., $\approx 2 \cdot
10^3$ or $\approx 3\cdot 10^2$, respectively).  The AES tree-based
search was performed with a 16-byte random key, and is using the Java
Crypto libraries to perform the actual AES calculations. The run time
for the tree-based search using AES was around 7 seconds, which
confirms the fact that AES requires significantly more calculations
than standard HB+ protocol. We can further estimate that since the
Meet-in-the-Middle PRF scheme takes about $\frac{2} {\sqrt{\beta}}=
0.063$ of the tree-based PRF scheme, it would take about 0.44 seconds
to run it on our computer. Therefore, our simulation shows that our
privacy-preserving method is faster then the Tree-based PRF and the
Meet-in-the Middle strategy. For $d=3$, our method is faster by a
factor of $\approx 53$ relative to the Tree-based PRF and $\approx 3$
comparing to the MITM method (for $d=2$, we get the factors of roughly
370 and 23, respectively).


\begin{table}
\caption{Simulation Results, $r=96$} 
\centering 
\begin{tabular}{|c| c| c| c| } 
\hline 
\textbf{Method}& \textbf{Comp. Time (seconds)}& \textbf{FAR}& \textbf{FRR}\\
		 & \textbf{mean$\pm$stdev}          & \textbf{expected/observed}
						&\textbf{expected/observed}
\\ [0.5ex] \hline\hline
ES (HB+) & 110.180~$\pm$~0.9656~~~ & 0.76~/~0.75 & 0.365~/~0.450 \\ \hline
Tree HB+ ($d=2$) & ~0.131~$\pm$~0.0234~ & 1.4E$-6$~/~0~~~~~~~~
		   & 0.432~/~0.422 \\ \hline
Tree HB+ ($d=3$) & ~~~0.019~$\pm$~6.75E$-4$ & 1.4E$-6$~/~2.14E$-6$
		   & 0.382~/~0.382\\ \hline
Tree PRF(AES) &    ~7.046~$\pm$~0.0561~ & --- &  --- \\ [1ex] 
\hline 
\end{tabular}
\label{table:SimRes80} 
\end{table}

The results of our simulation are shown in Table 5. The mean and
standard deviation of the run-time are provided, together with the FAR
and FRR for each case.
For the privacy-preserving Tree HB+ protocol, the FAR results were
not accurate due to the fact that the expected FAR is only $1.4 \cdot
10^{-6}$ and therefore a very large number of runs are needed to get
accurate results.
\section{Conclusions and Future Work} \label{conclusions.sec}

In this paper we developed tree-based HB-type protocols for
privacy-preserving authentication.  This is useful for two
reasons. First, very low-cost tags may be incapable of computing
standard PRFs and HB-type protocols are currently the only viable
solution for such tags. Second, since the underlying computations in
HB protocols are very efficient, it automatically reduces reader load
compared to PRF-based protocols.

We proposed some significant improvements over the naive use of HB+ in
a tree-based scheme. These improvements reduce the computation and
communication by a factor
of~$2\times$ to $4\times$. In fact our tree protocol is almost as
efficient as the underlying HB+.  The error
rates in our protocols are nearly as low as that underlying HB+
protocol. This makes our scheme suitable for a system with large
number of tags (Unlike exhaustive search, which would produce
unacceptable security for practical systems). We presented analytical
and simulation results comparing our method with prior proposals in
terms of computation, communication and memory overheads.

\end{document}